\documentclass[manuscript]{aastex}

\slugcomment{To be submitted to Ap. J. Lett., 45.}

\shorttitle{ Lateral Motion of Penumbral Filaments}
\shortauthors{Gosain et al.}

\begin{document}

\title{HINODE Observations of Coherent Lateral Motion of Penumbral Filaments during a X-class Flare}

\author{S. Gosain, P. Venkatakrishnan and Sanjiv Kumar Tiwari}
\altaffiltext{ }{Udaipur Solar Observatory, Physical Research Laboratory, P. Box 198, Dewali, Badi Road, Udaipur 313001, Rajasthan, India.}

\begin{abstract}
The X-3.4 class flare of 13 December 2006 was observed with a high cadence of 2 minutes at 0.2 arc-sec resolution by HINODE/SOT FG instrument. The flare  ribbons could be seen in G-band images also. A careful analysis of these observations after proper registration of images show  flare related changes in penumbral filaments of the associated sunspot, for the first time. The observations of sunspot deformation, decay of penumbral area and changes in magnetic flux during large flares have been reported earlier in the literature. In this {\it Letter}, we report lateral motion of the penumbral filaments in a sheared region of the $\delta$-sunspot during the X-class flare. Such shifts have not been seen earlier. The lateral motion occurs in two phases, (i) motion before the flare ribbons move across the penumbral filaments and (ii) motion afterwards. The former motion is directed away from expanding flare ribbons and lasts for about four minutes.  The latter motion is directed in the opposite direction and lasts for more than forty minutes. Further, we locate a patch in adjacent opposite polarity spot moving in opposite direction to the penumbral filaments. Together these patches represent conjugate foot-points on either side of the polarity inversion line (PIL), moving towards each other. This converging motion could be interpreted as shrinkage of field lines.
\end{abstract}

\section{Introduction}
Photospheric changes during large flares have been studied for a long time \citep{Giovanelli40,Howard63,Rust76,Tanaka78}. Such changes have been observed successfully in various parameters like, shear angle \citep{Ambastha93,Wang94,Hagyard99}, sunspot morphology \citep{Anwar93}, magnetic flux \citep{Lara00, Spirock02, Wang02a, Wang02b, Sudol05, Zharkova2005}, penumbral size \citep{Wang04,Deng05, Liu05}. Most of these observed changes are located in the penumbra of the sunspot. \cite{Hagyard99} argued that penumbral regions contribute most to magnetic energy of a force-free field (Low 1985) given by $E_M= \int{(x B_x + y B_y)B_z dx dy}$. Since $B_z=0$ near Polarity Inversion Line (PIL) and $B_T = \sqrt{B_x^2 + B_y^2} \approx 0$ in the umbra, these regions contribute little to magnetic energy of the force-free field. Since, flares are driven by magnetic energy changes, the most promising locations to look for flare related changes are the regions between umbra and PIL. This could explain why most of the flare related studies show changes in the penumbra.

\cite{Hagyard99} found interesting coherent variation of the shear angle in patches of the magnetic field on timescales shorter than the photospheric Alfven travel time over the patches. They had conjectured that electrodynamic effects of coronal magnetic field variations could induce coherent magnetic variations even at the photospheric levels. They also concluded that instruments with greater sensitivity could throw more light on these induced field variations. The Solar Optical Telescope (SOT; \cite{Tsuneta08}) onboard Japanese space mission {\it Hinode} \citep{Kosugi07} is one such instrument capable of detecting these changes.

The X-3.4 class flare, observed in NOAA 10930 on 13 December 2006 during 02:00 UT to 03:00 UT, by {\it Hinode} spacecraft, provided such an opportunity.  Various aspects of this active region have been reported. Three dimensional magnetic field reconstruction of this region was performed by \cite{Schrijver08} using  NLFFF extrapolation methods. They found regions of strong vertical current density between the two spots which  decreased significantly after the flare. Further, the free energy also decreased after the flare. The variation  of  the  weighted mean shear and total magnetic shear \citep{Wang08}  with altitude was studied using extrapolated fields by \cite{Jing08}. They found that near the PIL, up to a height of  $\approx 8 Mm$ the shear increases after the flare and decreases beyond that  up to $\approx 70 Mm$. Beyond 70 Mm the field remains close to potential before and after the flare.  However, the cadence of these vector magnetograms is insufficient to search for similar flare related changes as observed by \cite{Hagyard99}.

 In order to detect flare related changes in  magnetized regions like sunspots one can also use high angular resolution images obtained at a high-cadence,  during the flare interval. Such observations are nowadays routinely available from Filtergraph (FG; \cite{Tsuneta08}) instrument onboard {\it Hinode} space mission. The high resolution images allow us to investigate the evolution of the fine structure like penumbral filaments during a flare, e.g lateral motion and/or twist. In this {\it Letter}, we use  FG observations from {\it Hinode} to investigate  flare-induced changes during a X-class flare.

In section 2.1 we describe the observations while the method of registration applied on the data to establish the coherent motions are described in section 2.2. We present our results in section 3 while the discussion and conclusions are given in section 4.

\section{Observations and Data Analysis}
\subsection{Observational Data}
The high-resolution filtergrams of a flaring $\delta$-sunspot in NOAA 10930 were obtained by  Solar Optical Telescope (SOT) using filter-graph  (FG) instrument onboard HINODE spacecraft.  These  filtergrams are obtained in a 10 \AA ~band centered around 4305 \AA ~called G-band and in Fe I 6302.5 \AA at a cadence of 2 minutes. The morphological features seen in these wavelengths are believed to outline the magnetic structure of the penumbra in photosphere. The HINODE data are calibrated using standard libraries in the SolarSoft package. The  spatial sampling is 0.1  and 0.16 arc-sec/pixel for the G-band and Fe I 6302.5 \AA ~filtergrams, respectively.

The vector magnetic field, derived by Stokes inversion of the spectra obtained by spectro-polarimeter (Hinode/SP) instrument \citep{Lites07} under Milne-Eddington approximation \citep{Skumanich87}, are obtained from  \citet{Schrijver08}. SP scans during 20:30 UT on 12 December 2006 and 04:30 UT on 13 December 2006 are used.  These vector magnetograms  are resolved for 180$^\circ$ azimuthal ambiguity using ``minimum energy" algorithm \citep{Metcalf94} and are transformed to heliographic coordinates. Also, the vector magnetograms are rebinned and co-aligned with the FG filtergrams by manually registering the spatial features.

\subsection{Methods of Analysis}
The analysis of the high-resolution penumbral features during X-class flare was done using  a time sequence of registered G-band filtergrams. The registration of the filtergrams was done as follows:
(i) A time sequence, from 02:00  to 03:00 UT, spanning the  X-class flare, was selected for registration.
(ii) A large portion of the field-of-view (FOV), as shown in top-left panel of figure 1, was extracted from the full frames. The first image at 02:00 UT was then used as a reference image and the next image at 02:02 UT was registered  with respect to  the reference image using cross-correlation technique and sub-pixel interpolation \citep{November1988}. This registered image acts as the reference frame for the next image at 02:04 UT. This process was repeated until the entire time sequence was co-aligned. Small scale features evolve rapidly within the 2 minute time interval of the consecutive frames. This results in a decrease of correlation and increases the error in registration. On the other hand, a larger field-of-view will allow proper registration of the larger scale features like sunspots, which do not significantly evolve over 2 minutes.
Registration using larger features keeps the sunspot on the whole globally co-aligned, thereby allowing us to detect the relative motion of the small scale features.

Figure 1(a) shows the G-band field-of-view that was registered. The regions of interest that are discussed in the section 3 are marked by white rectangles and numbered as `1' , `2' and `3'.  Figure 1(b) shows the running difference between frames obtained at 02:32 and 02:24 UT. Figure 1(c) and 1(d) show the longitudinal component of the SOT/SP vector magnetic field, obtained at 20:30 UT on 12-December-06 and at 04:30 UT on 13-December-06 respectively. These longitudinal field maps are co-aligned with the G-band field-of-view in figure 1(a). The spatial average of magnetic field parameters inside the box `1' are given in Table 1.

In order to validate the registration, we take the running  difference of the images. Any shift due to improper registration or spacecraft pointing errors will appear on smaller scales as dipolar black and white features with constant orientation of the dipole all over the difference image. Further, any rotation of the entire field-of-view (e.g. caused by spacecraft rotation) would lead to an antisymmetric pattern of dipoles. The evolution of the features will result in a random orientation of the dipolar features which can be distinguished from the afore-mentioned systematic patterns.

Figure 2(a) shows the difference between images obtained at  02:00 and 02:04 UT, respectively. This corresponds to an epoch before the flare. Figure 2(b) shows the difference between images obtained at  03:52 and 03:56 UT, respectively. This corresponds to an epoch after the flare. We choose 4 minute gap to enhance the running difference signal. The absence of systematic patterns of black and white dipolar features confirms the registration. The absence of a systematic pattern at the umbra-penumbra boundary again confirms the registration.

We have made movies of the registered G-band images (available online). To confirm that the  lateral motion seen in the movies is not an effect of G-band intensity variation during flare, we also registered Stokes-V images taken in Fe I 6302.5 \AA\ and found a similar pattern of lateral shifts. The movies of Stokes-V images are also made available online. The flare ribbons are clearly seen as polarity reversals in the Fe I 6302.5 \AA\ Stokes-V movies, as reported earlier by \cite{Isobe07}.

Apart from the difference images,  we also made LCT velocity maps \citep{Welsch04} as shown in figure 2(e)-(h). These maps are made for the same timings as in figures 2(a)-(d), respectively. However, for LCT maps we use FGIV Stokes-V images in order to avoid false signals arising due to the brightness variations present in the G-band images.

We obtain the shifts $\Delta x$ and $\Delta y$ of features inside box `1' by cross-correlating consecutive frames with sub-pixel interpolation.
We derive the total horizontal displacement during the time interval 02:00 UT to 03:00 UT by integrating these shifts. The location of box `1' is chosen because this area shows large signal in the difference images and is the area of strongest motions, as seen in the movies.
Box `2', of same size as box `1',  is chosen for reference and is located far away from the neutral line. The X and Y displacement profile is shown in figure 3(a) and (b), with solid and dashed line corresponding to box `1' and `2' respectively. The velocity $v_x=\Delta x / \Delta t$ and $v_y=\Delta y /\Delta t$ is shown in figure 3(c) and (d). The GOES-10 soft X-ray flux is shown in figure 3(e).

\section{Results}

\subsection{\it Running difference of G-band images}
In figure 2(a) and (d), the difference signals are due to brightness variations along penumbral filaments caused by bright penumbral grains. In figure 2(b) and (c),  black-arrows are drawn to locate the G-band ribbons, while the rectangular boxes are drawn to show difference signal due to lateral shift of penumbral filaments.

It may be noticed  in figure 2(b) that: (i) the difference image has a dipolar intensity pattern with white on the right side of the dipole, (ii) difference signal is all along the filament suggesting coherent nature of shift rather than progressive nature, and (iii) even inside the white box, which is located much ahead of the ribbons, penumbral shifts can be seen. In figure 2(c) we notice that: (i) the difference image has a dipolar intensity pattern with black on the right side of the dipole, opposite to what is seen in figure 2(b). This suggests that the motion is in opposite direction after the ribbon has crossed.

\subsection{\it Local Correlation Tracking (LCT) Maps}
The flow vectors in figure 2(e)-(h) clearly show the following behaviour:
(i) during non-flaring times 2(e) and (h): the arrows point mainly along penumbral filaments,
(ii) during flare 2(f) and (g): the arrows point towards direction of lateral shift of the filaments. This shift is towards the right side in figure 2(f). The shift is towards the left side in figure 2(g).

\subsection{\it Horizontal displacements inside Boxes `1' and `2'}
In figure 3(a), the displacement is initially in the positive X direction and peaks at 02:22 UT. The displacement after 02:26 UT is in the negative X direction. A significant displacement in the positive Y-direction is seen after 02:26 UT in the figure 3(b).  The motion in box `2' does not show such behaviour. In figure 3(c) a sudden acceleration and then deceleration is seen during 02:20 to 02:28 UT. A similar behaviour for $v_y$ is seen in figure 3(d). The microwave observations in 2.6 to 3.8 GHz  peaks at around 02:22 UT  \citep{Zhang08}.  The first signatures of penumbral filament motion are seen at around 02:18 UT. The maximum shift in the X and Y direction is seen at 02:22 UT  i.e., coinciding with the peak in microwave emission.

The amount of energy E involved in the lateral motions occurring over a patch of area $A$  (64 arc-sec square, Box `1') during the flare (02:20 UT) of duration $t$ (240 seconds, 02:18 to 02:22 UT) is given by $E=\dot{E}t$, where $\dot{E}=\int{P}dx~dy$,  $P$ being the  Poynting flux given by $P=-\frac{1}{4\pi}(v_h \cdot B_h)B_z$ \citep{Ravindra2008}. To calculate this energy we need $B_h , v_h $ and $B_z$. Although we know that the magnetic field varies by several tens of percent during certain flares \citep{Zharkova2005}, the lack of high cadence vector magnetograms during this flare forces us to take average values for estimating the energy. From the spatial average of horizontal velocity during the flare  and from the average of  the vector field  before and after the flare  derived for box `1'  (column 4 of Table 1), we estimate the energy to be $\sim 1.45 \times 10^{30} ergs$. The sign of the energy flux is negative showing that the magnetic energy flows into the photosphere.  \cite{Schrijver08} reconstructed three-dimensional nonlinear force-free field (NLFFF) for this active region. They found  a reduction of free-energy of about $\sim 3 \times 10^{32}$ ergs, after the flare. The free energy released during the flare as estimated by (Schrijver et al 2008) is valid for entire volume of the active region, while our estimate is only for a local patch. The energy involved in the motion of penumbral filaments is thus only about a percent of the total change in the magnetic energy. Hence a small part of the energy released in the large scale coronal
changes can be available for causing the coherent lateral motion of the
penumbral filaments. Further, because the small scale lateral motion
starts before impulsive energy release in the flare, we could also
consider this as a manifestation of a precursor that trigger the large
scale changes leading to the CME.

From the analysis of difference images and LCT maps, as well as horizontal  displacements in box `1'  it is now clear that there are two phases of filament motion: (i) an impulsive motion directed away from the ribbons during 02:18 to 02:22 UT, and (ii) a long-lived motion after 02:22 UT in the reverse direction.

In the G-band movie the locations marked `1' and `2' correspond to conjugate foot-points, as inferred from the extrapolated field lines in \cite{Schrijver08}. These locations show  motion towards each other during 02:28 UT and 02:32 UT. The converging motion can be easily seen by toggling between these two frames in the movie.

\subsection{Evolution of magnetic twist parameters}
 We calculated the twist parameter $\alpha_g$ and mean Signed Shear Angle (SSA) for this sunspot using the methods described by \cite{Tiwari09a,Tiwari09b}. Both parameters represent global twist of the magnetic field; however mean SSA has the advantage of being useful even for non-force-free  fields. The value of $\alpha_g$ is $-9.4 \times 10^{-8} m^{-1}$ before the flare and $-7.3 \times 10^{-8} m^{-1}$ after the flare. The mean SSA is -9.5 and -8.6 degrees before and after the flare. Both  parameters are found to decrease after the flare which is consistent with decrease in free energy estimated by \cite{Schrijver08}.  Also, the results are consistent with the findings of \cite{Wang08} i.e., decrease of magnetic shear after the flare.

\section{ Discussion and Conclusions}
In this {\it Letter}, we present sub-arcsec (0.1"/pixel) observations in G-band at a cadence of 2 minutes, taken during 13 December 2006 X3.4 flare. High-resolution observations of the photosphere during a large X-class flare have never been done before with the kind of stability and image quality that HINODE/SOT provides. A careful analysis,  after proper registration of these images, show  flare related changes in penumbral filaments of the associated sunspot, for the first time, in the form of coherent lateral motions of the penumbral filaments.

We found the motion of the penumbral filaments to exist in two phases: (i) the first phase is short-lived (about 4 minutes) with the time of maximum displacement coinciding with the peak of the microwave flux, and (ii)  the second phase is of longer duration (more than 40 minutes).

Further, we observe the patches corresponding to conjugate foot-points in either spots moving towards each other. These conjugate foot-points, labeled as `1' and `2' and marked by arrows, are clearly seen in the online G-band movie. This converging motion of conjugate foot-points could be interpreted as shrinkage of field lines during flare, which is conjectured by \cite{Hudson00}. However, we lack high-cadence vector field measurements in this case and our present inferences are purely from G-band images. We have also shown, in section 3.4, a decrease in the global twist after the flare. It will be interesting to see the evolution of local twist as well as global twist during the flare. We expect to detect flare-related vector field changes with  high-cadence vector magnetograms from the upcoming SDO mission.

\cite{Schrijver08} found strong arching filamentary current system embedded in the magnetic flux emerging between the penumbrae of the two spots, prior to flare.  These currents are significantly reduced after the flare. This current system seems co-spatial with the field lines that connect the conjugate foot-points discussed above.

 Currents in astrophysical plasmas are produced by the distortions of the magnetic fields given by the curl of the field \citep{Parker79}. Generally these distortions are driven by plasma dynamics. The coherent nature of the distortions in the present observations, occurring within a time scale shorter than the Alfven travel time over the affected area, rules out the plasma dynamics of the photosphere as the cause of the distortions. It is more likely that the distortions are caused by the fields induced in the photosphere  in response to the restructuring  (like untwisting of the global field, as inferred from the decrease in SSA) of the coronal magnetic fields,  which are line tied to the photosphere. We have estimated the energy of the induced lateral motions of the penumbral filaments to be $\sim 1.45 \times 10^{30}$ ergs, which is a small fraction of the total magnetic energy of $ \sim 3 \times 10^{32}$ ergs released in the flare \citep{Schrijver08}. This mechanism of induced photospheric magnetic response to coronal magnetic restructuring appears energetically feasible to drive the lateral motion of the filaments. We call this first phase the ``rapid restructuring phase" (RRP).

The slow global change in the displacement of the filaments during second phase can then be considered as a response of the gradual restructuring of the field on the much longer Alfven time-scale corresponding to the larger size of the CME flux rope system \citep{Liu2008}. We call this phase the ``gradual restructuring phase" (GRP). However, we need more observations to confirm these phases in other flares.

Vector magnetic field maps at a cadence similar to G-band images are not available from SOT/SP. However, future  instruments, like HMI aboard SDO mission,  will have high-resolution vector magnetic field maps at high-cadence and will shed more light on the flare related vector field changes associated with such lateral motions of small-scale features.

\acknowledgements
We thank the referee for his/her valuable comments.  We thank Dr. Brian T. Welsch and Dr. C. J. Schrijver  for providing  LCT code and the vector magnetic field data, respectively. {\it Hinode} is a Japanese mission developed and launched by ISAS/JAXA, with NAOJ as a domestic partner and NASA and STFC (UK) as international partners. It is operated by these agencies in co-operation with ESA and NSC (Norway).

\begin{table}
\caption{Average values of magnetic and dynamical parameters in box '1'}
\label{tab:1}
\begin{tabular}{lccr}
\hline
Parameter &  Before Flare  &  After Flare & Mean Value \\
\hline
Field Strength (Gauss)  & 2072  & 2333  & 2202 \\
 Inclination (Degrees) & 133  & 131  & 132\\
 Azimuth (Degrees) & 72  & 67  & 69\\
 $B_x$ (Gauss)& 324  & 582 & 453\\
 $B_y$ (Gauss) & 1276 & 1493 & 1384 \\
 $B_z$ (Gauss) &  -1472& -1600  & -1536 \\
 $v_x$ (m/s) & $--$ & $--$ & 125\\
 $v_y$ (m/s) & $--$ & $--$ & -125\\
\hline
\end{tabular}
\end{table}

\begin{figure}
\begin{center}
\includegraphics[angle=0,scale=.60]{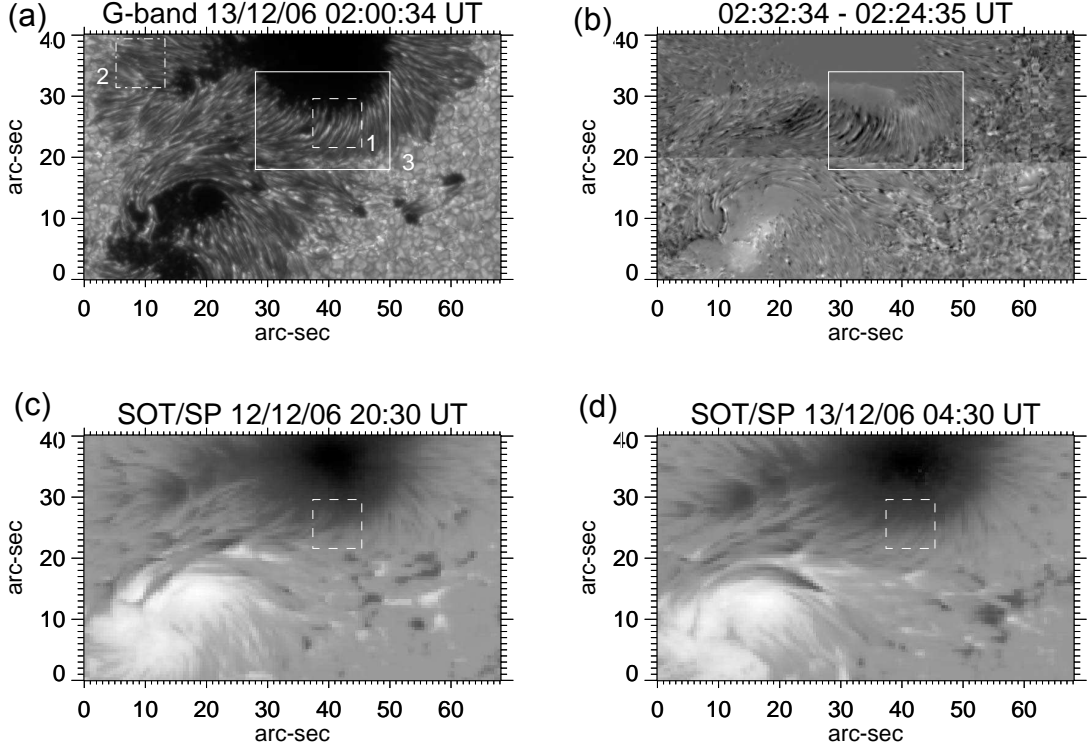}
\end{center}
\caption{ Panel (a) shows the field-of-view selected for the registration of G-band filtergrams. The box `1' and `2' are used to derive the total horizontal  displacement during 02:00 to 03:00 UT. Box `1' corresponds to region where penumbral filaments move in response to flare, while box '2' represents a reference region located away from the flaring region. Box '3' is used for showing running difference signals. Panel (b) shows a running difference image during flare. White box marks the location where clear signals of lateral motion are seen. Panels (c) and (d) show longitudinal field maps derived from SOT/SP scans during 20:30 UT on 12 December 2009 and 04:30 UT on 13 December 2009. These longitudinal maps are co-aligned with the G-band field-of-view shown in (a). The white box represents the location of box '1'. The spatial average of magnetic field parameters inside box `1' are given in Table 1. }
\end{figure}

\begin{figure}
\begin{center}
\includegraphics[angle=0,scale=.1]{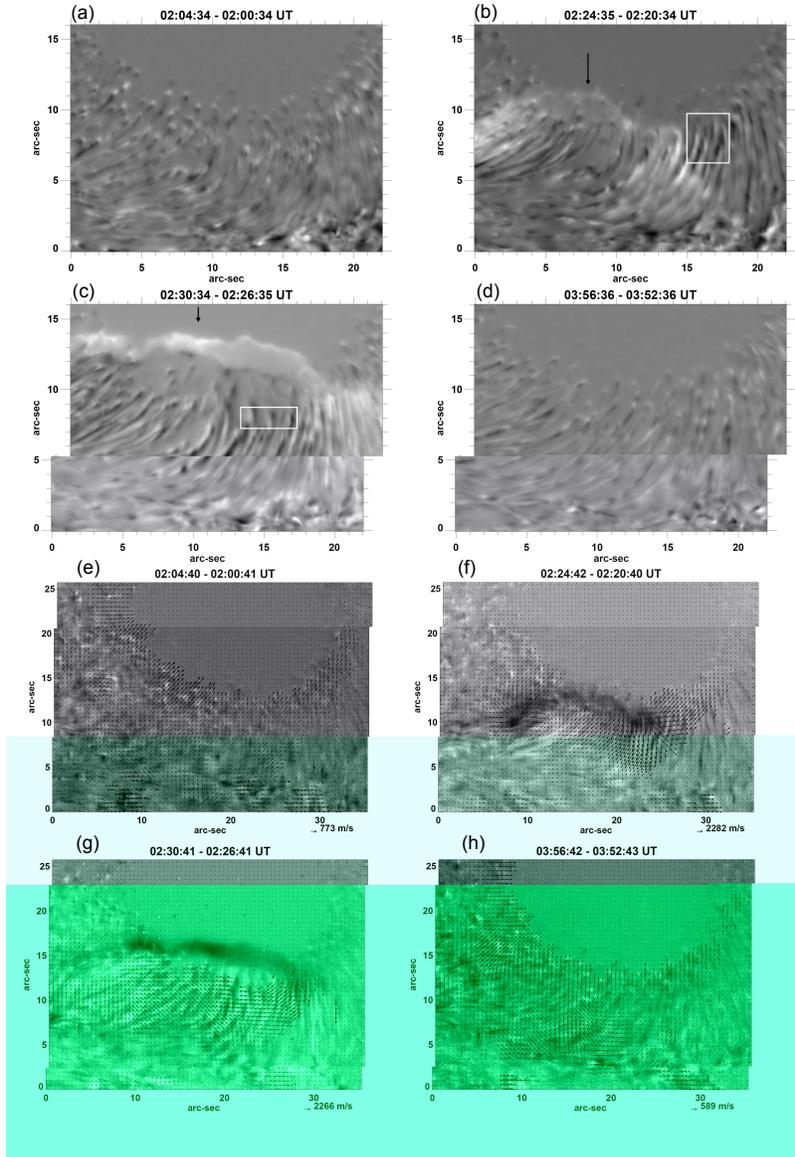}
\end{center}
\caption{Panels (a)-(d) show the running difference of G-band images corresponding to the box '3'.  The black arrows  indicate the flare ribbons. Panels (e)-(h) show horizontal flow vectors corresponding to FGIV Stokes-V images derived using LCT technique  (overlaid over the FGIV running difference images).}
\end{figure}

\begin{figure}
\begin{center}
\includegraphics[angle=0,scale=.8]{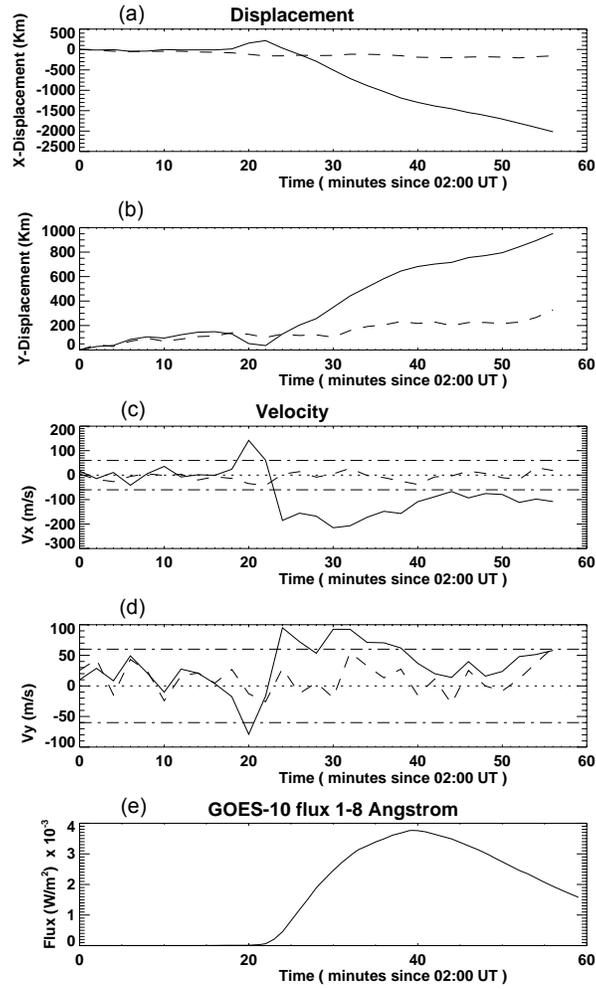}
\end{center}
\caption{The panels (a)-(e)  show X and Y displacement, X and Y velocity and GOES-10 soft x-ray flux during 02:00 to 03:00 UT. The solid and dashed curve correspond to Box '1' and '2' respectively.  The dash-dotted horizontal lines in third and fourth panel represent velocity corresponding to a shift of $\pm$ 0.1 pixel.  }
\end{figure}



\end{document}